# Effort minimization in UI development by reusing existing DGML based UI design for qualitative software development

P. K. Suri
Professor, Department of Computer Sc. & Application
Kurukshetra University, Kurukshetra,
Haryana, India
.

Gurdev Singh
Department of Computer Sc. & Application
Kurukshetra University, Kurukshetra,
Haryana, India
.

*Abstract*— **This paper addresses the methodology for achieving the user interface design reusability of a qualitative software system and effort minimization by applying the inference on the stored design documents. The pictorial design documents are stored in a special format in the form of keyword text [DGML tag based design]. The design document storage mechanism will expose the keywords per design stored. This methodology is having an inference engine. Inference mechanism search for the requirements and find the match for them in the available design repository. A match found will success in reusing it after checking the quality parameters of the found design module in the result set. DGML notations produces qualitative designs which helps in minimizing the efforts of software development life cycle.**

## I. INTRODUCTION

A new industry practice emerged in quality software development which deals with not only developing a run away solution for a problem but to get a better solution which could be reused. Generally, we spend a lot of our time and efforts in developing solutions for difficult, time consuming and mission critical problems. One of these solution set components is the UI designs document [1][2]. Design process takes maximum time and efforts and is never reused in the future.

Reusability is achieved up to some level in coding practices like OOPs, Component based developments, Active-X, technology where a piece of written code is reused after passing few checks for non discloser of the blueprints of the solution [4].

We propose that the root of reusability is there at the starting and bridging activities of the software development process that is design phase. This includes functional and user interaction design. We can consider the reusability at this level, such as UI design reusability [6][14]. If a UI design is made reusable, we can achieve the reusability in the development phases because this coding phase is very much driven from the software design.

This paper proposes a new approach in the context of the User interface design reusability. Reusable UI design methodology is of special concern with the effort minimization, which could be achieved by following our approach. This paper also discusses the notation of the design document storage for reusability.

Reusable UI design approach by inference is having the full impact of the strategic fitness in achieving the large and tough goals in small time and smartly. This approach will fit as getting a good domain solution in minimum efforts. A good review mechanism can also be imposed on the stored reusable UI design for assigning quality attributes.

Proposed approach eliminates the need for fresh efforts for UI design every time a solution is required. The UI design document having diagrams, images will be stored in textual format and every design stored will have a few keyword and properties [3][5]. Keyword submission per design facilitates the search against the requirement specifications and outcome of the result having the attribute tag of quality benchmark, like number of times the design is reused. The research also includes one tool, which helps in maintaining the central repository of the different design documents and inference mechanism on the stored design repository with in the organization.

## II. DGML BASED UI DESIGN REPRESENTATION

The reusability of user interface design for a software system could be achieved by storing the graphical design documents in the form of text. The graphics-symbols of the design document can be represented in the form of text. Tools are available which allow the user to draw the user interface with diagrams. These tools however lack one aspect to store the design elements in the form of structured text as backend.

Our approach is to device some mechanism using which we can design the diagrams for user interface and represent them in the text format. So that when user creates one UI design and save it, it generates one text file also. The approach that we choose is to represent the text per design element is as a XML tag. So every design element will have a unique tag.

Initially the process start with making a new UI design document, which is XML, based. The system to create the UI design is having the collection of UI elements. All finite design elements have assigned names. User has to select and place these design elements while creating design document. When one UI design module gets completed, there will be one complete XML document associated with it. These UI design repository will be a collection of associated XML document per design as shown in following figure2.

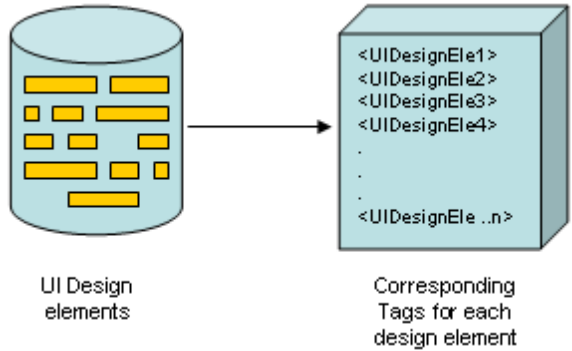

Figure 1: Every UI design element is having UI design element tag associated with it.

There will be a repository of designs after following this approach and every design is well defined in terms of XML tags. The text format of storing the design is having the predefined tags. Text format will be stored in XML notations. We say this as Design Markup Language and are having the tags for all knows design elements. When user picks one design element, its respective tag will be placed in its corresponding text file. When UI design complete for some scenario by following this approach, system will be having a complete representation in textual format in the form of tags.

The XML text file having the details of design figures is the basis of our research. The approach has many benefits. When pictorial representations are saved in text form, lots of possibilities are there which increase the overall development process [10][12]. Reusability of design search with in the available UI design, automatic generation of design skeleton on the basis of requirement document and available design text keywords, auto generation of test case scenarios.

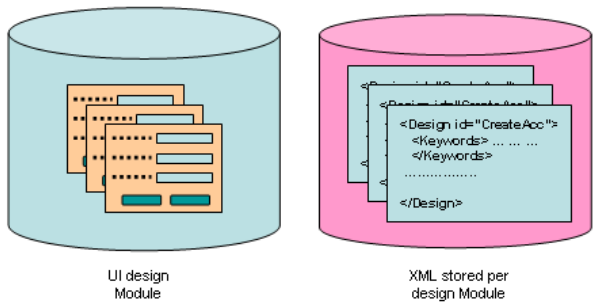

Figure 2: Each design module is having associated DGML document.

Along with the design element tags, the XML-DGML file is having the information with the help of special tags (like <keywords>), which is important and is a key for further reusability criteria implementation. This additional information is the list of keywords that designer wants to assign to his design. Each UI design is having one name, keywords and some attributes. Name identifies the design module, keywords are the handles for reusability, and attributes are factors, which design document gain after reusability. Experts review the UI design and assign scores. A good score by expert make design module a good candidate for reusability.

When some user wants to create the new design, he has to submit the UI requirement specification. The requirement specification document is analyzed and some keywords are evaluated from this. These are requirement specification keywords. Next step is that system looks into the available keywords of design module repository again these keywords. This special search process to find out required design from existing design is the design inference

The design repository will be stored for each design element in a centralized database of the organization. The design stored is of the atomic nature by representation of text format. Each elementary design element will be having one name and some associated keywords. These keywords will be reflected in the centralized repository.

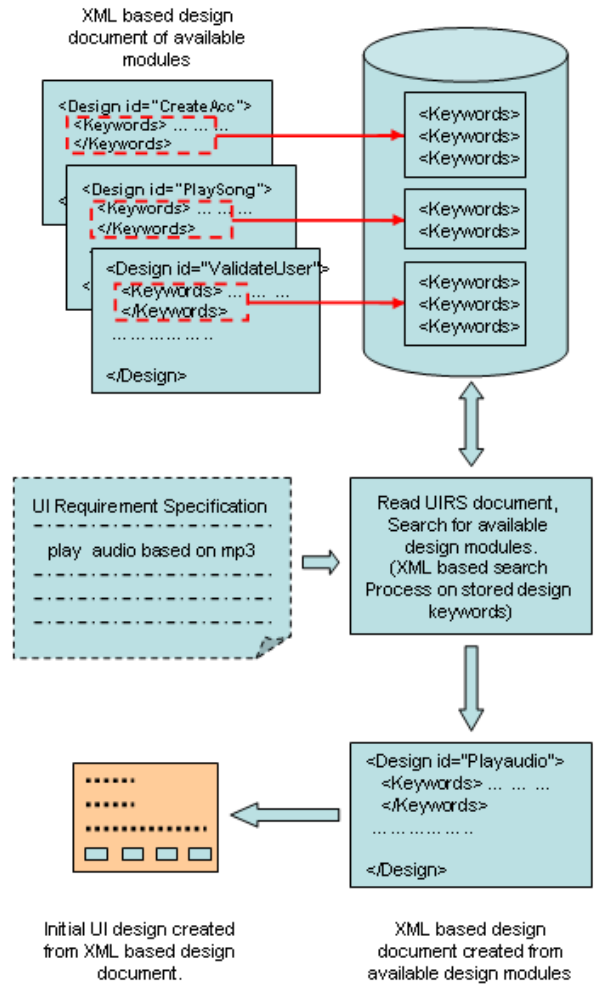

Figure 2: Mechanism for UI design reusability.

In software development life cycle, when analysis phase is completed and the requirements are defined, next stage is to move to design process. The existing design repository will help out in reusing the design elements.

Above figure shows the mechanism for UI design reusability. After following the approach for creating XML based UI design documents (DDML), we will be having a database of XML sheets for each design module. Each design module is having the name and keywords associated with it.

*< !ELEMENT name *>*

*< !ELEMENT keywords *>*

Inference will be preformed on the centralized design repository against the requirement specification keywords from the UIRS (User interface requirement specifications). This will be a type of look-in search process in the design repository keywords of elementary design. When search gets completed, it will generate a minimum UI design document having the maximum reusability of existing modules. This DGML document is arranged in order to get the maximum requirement fulfillment for required UI design.

Also, the result of the inference on stored design repository will be having the elementary design modules having attributes also. These attributes specify the design reusability factor for a module (DRF). More the reusability factor of a module indicates that module is strong candidate for reusability in new design. Reusing a UI design module will increase the reusability count factor by one.

### III. WORKING

*A.) Create DGML based UI design and generate design repository per design module.*

Figure 4 is the logical representation while creation of new design. If user creates the new design and it is driven from some existing design element Dg..n, it will increment the design reusable factor DRF for Dg..n by one. This increment in DRF make makes design Dg..n a stronger candidate for further reusability

*B.) Retrieve DGML based design repository for reusability and create initial proposed design layout.*

The following flow chart (Fig 4) shows how the DRF (Design Reusable Factor) helps in reusing the existing UI design. First the search for the keywords in existing design will be performed. The outcome then will be sorted in descending order on the bases in DRF. The top value of result outcome shows that first one in the search is the best candidate for reusability

### IV. EXPERIMENTAL DETAILS

For the experimental validation of the proposed approach of UI design reusability and effort minimization, we have considered five small projects having very small UI requirements. These projects are having maximum of six different modules. UI design requirement for these modules have been studied and we also consider one end user for our experiment to which we will give the finished design for review. (We assume and prepare this person with all the knowledge of the UI design, like its functionality scenario etc so that he may evaluate the UI design).

From requirement discussion to first draft, we record the time spend for each project. We also record the time consumed by the UI design architect. When this gets finished, we need to discuss it with the end user. Time record for the end user involvement was also recorded. We include the end user in the design phase for layout, look and feel type observation. We consider this as design prototype.

Table1 is the recorded of the efforts made by the UI design architect and end user for verification of the initial draft.

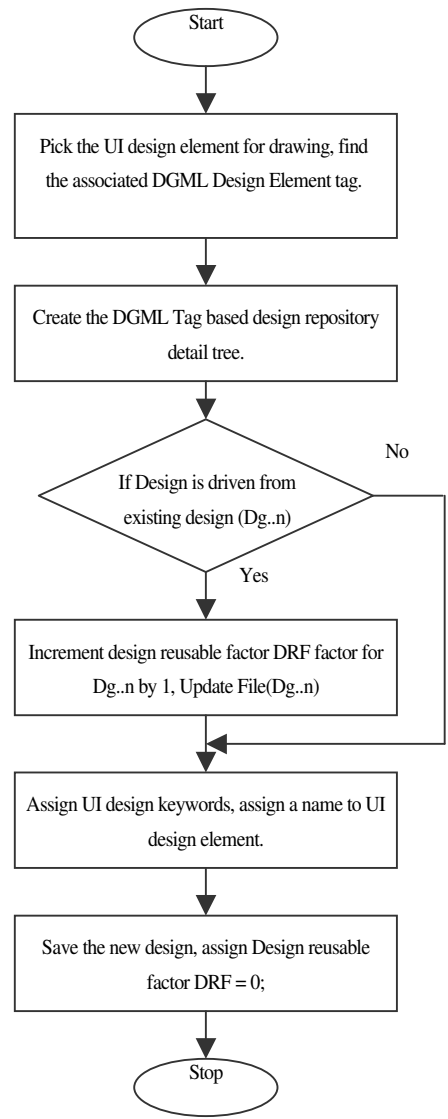

Figure 4: Create DGML based UI design

This is the observation of the projects using conventional system and we will compare it with efforts consumed using our approach in next section with Table 2.

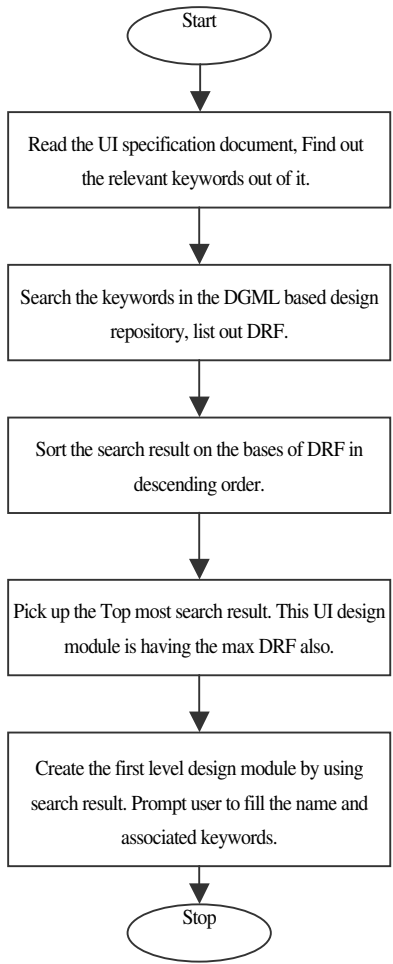

Figure 5: Retrieve DGML based design repository for UI design reusability

Table 1: Efforts (in Hrs) in UI design using conventional design methods.

| Modules | Efforts in Hrs | |
|---|---|---|
| | Conventional UI Design efforts | End User involvement |
| Project1-4Modules | 13 | 6 |
| Project2-4Modules | 29 | 8 |
| Project3-5Modules | 17 | 7 |
| Project4-5Modules | 8 | 4 |
| Project5-5Modules | 19 | 7 |
| Project6-6Modules | 20 | 7 |

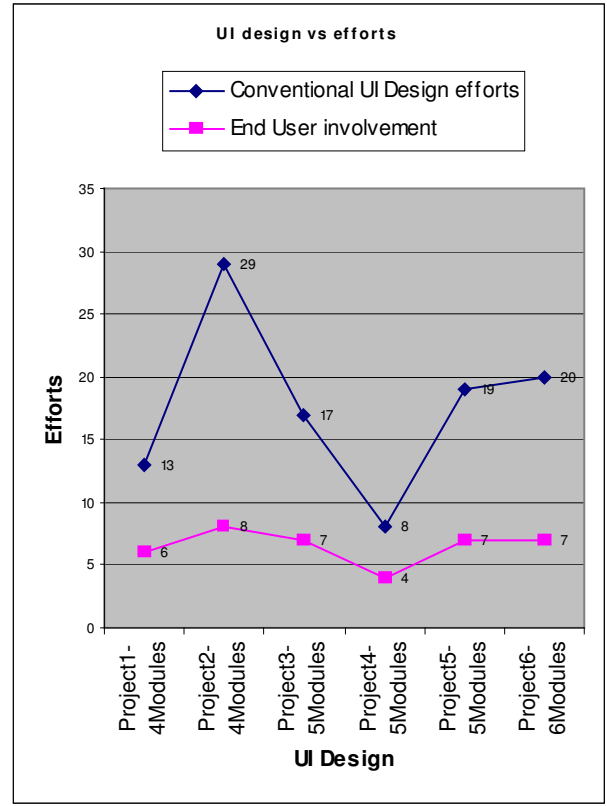

Figure 6: Efforts (in Hrs) in UI design using conventional design methods (Source: Table1).

The graph (Figure 6) represents that for a project of 4 modules, minimum time for design is 13 hrs where as maximum could be 29 hrs for a project of 4 modules. This variation shows that module count is immaterial but the important is nature of project. This observation when compared with the DGML based approach will shows very interesting results.

For our experiment with DGML based UI design, we are having 10 designs in the DGML storage and these are having 50% resemblance with the domain of the requirement. These design documents are having the keywords, which could be used during the search process.

The DGML based design experiment for the reusability of existing design requires the framing of the requirement specification in some special format. So, we have made the requirement specification document. This is having the keyword-based text instead of the plain English based scenario discussions. Same projects with special formatting of the requirement specification are submitted to DGML inference engine. This take very less time in identifying the candidate from available stored design and displays the results. The time taken in framing the requirement and getting the first level design from existing designs is shown in the following table (Table 2). Its approximately 1hr because this is not simply selection process, user have to spend some time (in few min) to identify which one is appropriate on the bases of some ranks, like reusability factor.

Table 2: Efforts (in Hrs) in UI design using DGML based UI design methods.

| Modules | Efforts in Hrs ||||
|---|---|---|---|---|
| | Efforts in Framing Requirements | Effort in generating first level design | Agile client efforts | Total efforts in automation |
| Project1-4Modules | 7 | 1 | 1 | 9 |
| Project2-4Modules | 10 | 1 | 2 | 13 |
| Project3-5Modules | 7 | 1 | 2 | 10 |
| Project4-5Modules | 3 | 1 | 1 | 5 |
| Project5-5Modules | 5 | 1 | 3 | 9 |
| Project6-6Modules | 7 | 1 | 2 | 10 |

Observations in Table2 are showing the improvement in the efforts consumed. The only efforts consumed are in the framing of the requirement in the special format.

The agile client efforts, which can be a part of team also need to spend small time as compared to time spend in table1 by the user [15][16]. We propose the involvement of the user in the design process with our mechanism. User interactions could be of the type of selecting one design layout out of available design layout produced by the system.

The total time consumed in the finalization which is summation of time consumed in framing requirement, design generation by system and client time to select. We represent this as TED, i.e. total effort required for DGML based design.

$$TED = \sum_{i \in SP} RFi \oplus DGi \oplus ACEi$$

Here RF is the efforts required for requirement framing for software project SP, DG is efforts in design generation and ACE is the agile client efforts. Figure 6 shows the graphical representation for the same.

As an observation from figure7 we can see the total efforts consumed in the proposed design are less than that of conventional system which is showing the effort minimization after using the DGML based design approach and then applying the DNSIM (Design notation storage and inference mechanism) for reusing the same for achieving the qualitative designs.

Table 3: Total Efforts comparison in using two approaches

| Modules | Total efforts in DGML based approach | Total efforts in conventional UI design. |
|---|---|---|
| Project1-4Modules | 9 | 19 |
| Project2-4Modules | 13 | 37 |
| Project3-5Modules | 10 | 24 |
| Project4-5Modules | 5 | 12 |
| Project5-5Modules | 9 | 26 |
| Project6-6Modules | 10 | 27 |

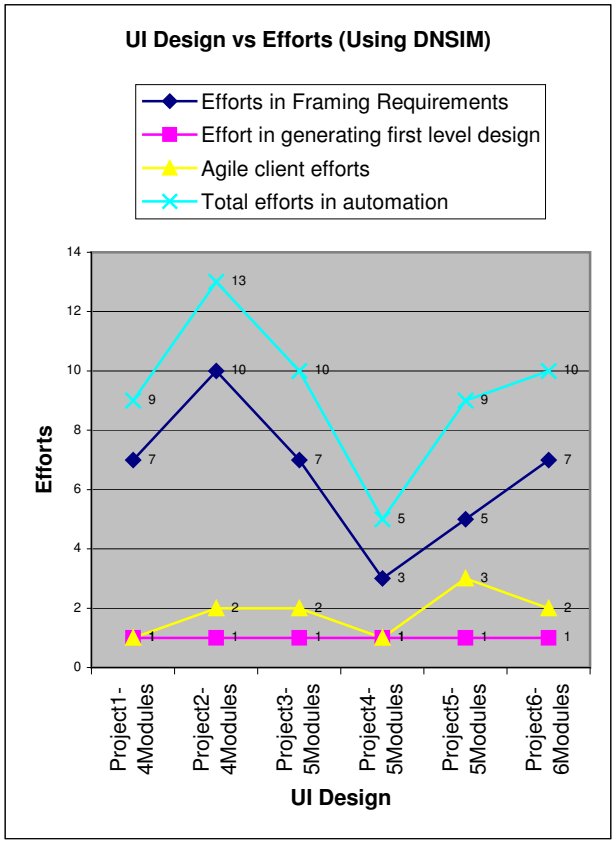

Figure 7: Efforts (in Hrs) in UI design using conventional design methods (Source: Table2).

Graphs below (figure 8) shows the comparison and figures in the total effort minimization. DGML based design approach uses requires the less efforts as compared to other methods. DGML based approach produces qualitative design, which helps in minimizing overall efforts in software development life cycle.

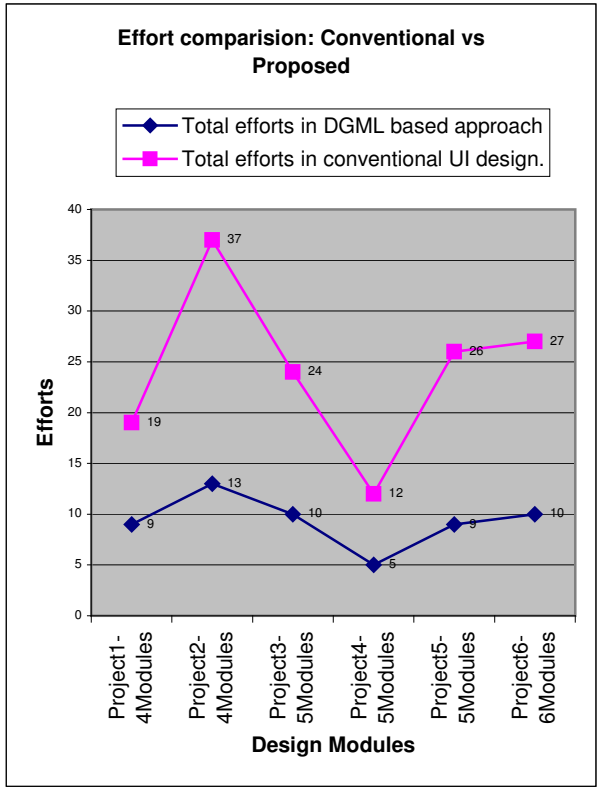

Figure 8: Total Efforts comparison: DGML based approach and conventional approach (Source: Table3).

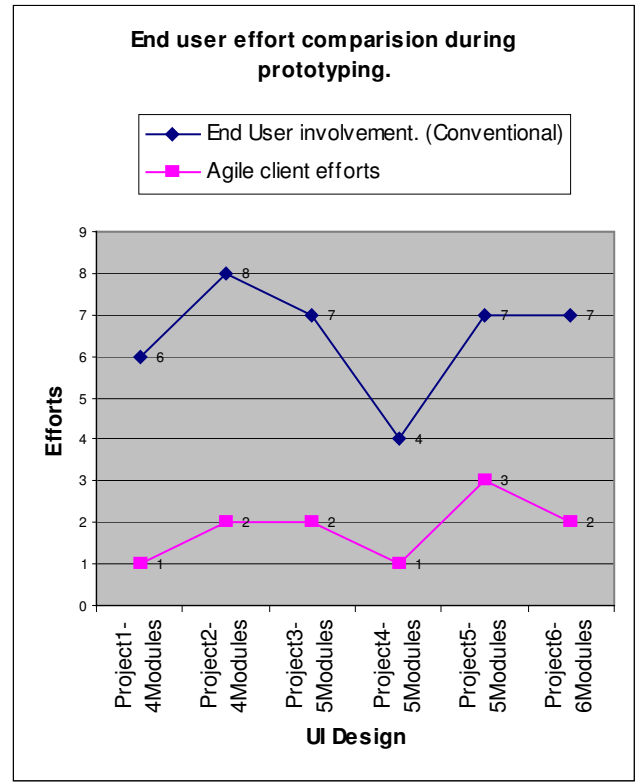

Figure 9: Total Efforts comparison: DGML based approach and conventional approach (Source: Table3).

The end user involvement and efforts for feedback are comparatively less in case of the DGML based approach. For same projects taken in our experiment, we found that there is a big difference in the end user involvement. As for project1 having 4 modules, the efforts consumed by the agile client for feedback are 6hrs for conventional approach where as its only 1hr required for the DGML based user interface design. This also represents that it's a qualitative design approach towards effort minimization and making SDLC less complex.

## V. DISCUSSION

DGML based designs representation is a new approach for storing the design elements in the form of specially design DGML-XML tags. These are plain text files, so we can write programs to manipulate and analyze these files for further enhancement.

To simulate the effort minimization, we carry out above-mentioned experiment. It shows that the there is tremendous decrease in the user involvement. We carry out our experiment on five different projects having different number of modules. Table1 is showing the efforts in man hrs and end user involvement using the conventional, approaches of software design engineering. The graph in the figure6 depicts that there is big involvement of the end user in the design phase (we consider agile methodology where customer is a member of development team) as for project2 29 hr efforts are from designer team and end user spend 8 hrs.

Table 4: User involvement efforts comparison in using two approaches.

| Modules | End User involvement. (Conventional) | Agile client efforts |
|---|---|---|
| Project1-4Modules | 6 | 1 |
| Project2-4Modules | 8 | 2 |
| Project3-5Modules | 7 | 2 |
| Project4-5Modules | 4 | 1 |
| Project5-5Modules | 7 | 3 |
| Project6-6Modules | 7 | 2 |

Table 2 shows the overall efforts for creating the first level design using DGML based approach. The total efforts in automation are less than that of the manual and conventional system. There is a significant minimization in the efforts of the agile client involvement as depicted from the figure7 and table2.

Figure8 shows the comparison between the total efforts of DGML based design and conventional design. In DGML based design, there is only on activity, which consumes time and that is the framing of the requirement specification document. This contains finding out the important keywords from the requirement specification document and using them later in the search process DGML based parsing.

Figure9 compares the involvement of the end user in the same projects using the two different approaches. It clearly shows that agile client involvement is reduced significantly after following the DGML based approach.

VI. CONCLUSION

The user interface design reusability leads to faster development of the application especially in the area where the UI is of prime importance. Big time and efforts, which are consumed in developing the core user interface design functionality, could be minimized. The user involvement in the design phase and efforts could be minimized in the design process. This is because the design elements are stored in the textual format and we can apply several kinds of search algorithms for finding the best solution for the problem. Further, design storage in specified format makes the system scalable and maintainable. The approach of the reusability of software design is novel. This paper is an initiative towards using DGML based UI design representation for reusability. The experiment carried on five projects for effort minimization using UI design reusability is showing interesting and valuable observation of the concept.

## AUTHOR INFORMATION

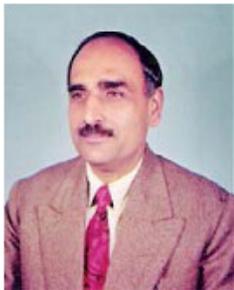

Dr. P.K. Suri received his Ph.D. degree from Faculty of Engineering, Kurukshetra University, Kurukshetra, India and master's degree from Indian Institute of Technology, Roorkee (formerly known as Roorkee University), India. He is working as Professor in the Department of Computer Science and Applications, Kurukshetra University, Kurukshetra–136119 (Haryana), India since Oct. 1993. He has earlier worked as Reader, Computer Sc. & Applications, at Bhopal University, Bhopal from 1985-90. He has supervised twelve Ph.D.'s in Computer Science and eleven students are working under his supervision. He has more than 125 publications in International/National Journals and Conferences. He is recipient of 'THE GEORGE OOMAN MEMORIAL PRIZE' for the year 1991-92 and a RESEARCH AWARD –"The Certificate of Merit – 2000"for the paper entitled ESMD – An Expert System for Medical Diagnosis from INSTITUTION OF ENGINEERS, INDIA. His teaching and research activities include Simulation and Modeling, Software Risk Management, Software Reliability, Software testing & Software Engineering processes, Temporal Databases, Ad hoc Networks, Grid Computing, and Biomechanics.

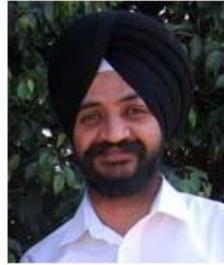

Gurdev Singh received his Masters degree in Computer Science from Department of Computer Science & Applications, Kurukshetra University, Kurukshetra, Haryana, India. Since 2002 he is working as Software Development Professional and had experience of working with MediaTek, and Siemens Information System, India. Currently he is working as senior software engineer for Samsung Electronics in Noida, India. He has completed projects in the field of software development for mobile devices. He loves to transfer user requirements in to piece of software. His interest includes work in the domain of software engineering, effort minimization in software development, qualitative software design and synchronization, software design representation methodologies and reusable software design techniques. He has written many papers in the related domain. He is fond of studying about the digital electronics and experimenting the same.